\documentstyle[bo99,epsfig]{article}

\title{The ASCA Hard Serendipitous Survey (HSS): a Progress Update }

\author{R.~Della Ceca$^1$, V.~Braito$^2$, I.~Cagnoni$^3$, T.~Maccacaro$^1$}

\affil{1) Osservatorio Astronomico di Brera, Milan, Italy;
       2) Universit\`a degli Studi di Milano , Milan, Italy;
       3) International School for Advanced Studies, SISSA, Trieste, Italy.}

\begin{document}

\maketitle

\begin{abstract}

We present here a status update on the ASCA Hard Serendipitous Survey
(HSS), a survey program conducted in the 2-10 keV energy band.  In
particular we discuss the number-flux relationship, the 2-10 keV
spectral properties of the sources and of the spectroscopically
identified objects.

\end{abstract}

\section{Introduction}

At the {\it Osservatorio Astronomico di Brera} we started a few
years ago the ASCA Hard Serendipitous Survey (HSS): a systematic search
for sources in the $2-10$ keV energy band, using data from the GIS2
instrument onboard the ASCA satellite.
The specific aims of this project are: a) to extend to faint fluxes the
census of the X-ray sources shining in the hard X-ray sky, b) to
evaluate the contribution to the Cosmic X-ray Background (CXB) from the
different classes of X-ray sources, and c) to test the Unification Model
for AGNs.

This effort has lead to a pilot sample of 60 sources that has been used
to extend the description of the number-counts relationship down to a
flux limit of $\sim 6\times 10^{-14}$ erg cm$^{-2}$ s$^{-1}$ (the
faintest detectable flux) resolving {\it directly} about 27\% of the (2
- 10 keV) Cosmic X-ray Background (CXB), and to investigate their X-ray
spectral properties (Cagnoni, Della Ceca and Maccacaro, 1998; Della
Ceca et al., 1999).

Recently the ASCA HSS has been extended: we discuss here this extension and 
the main results obtained so far. 


\section{The ASCA HSS Sample}

The data considered for the extension of the ASCA HSS were extracted
from the public archive of 1629 ASCA fields (as of December 18, 1997).
The fields selection criteria, the data preparation and analysis,
the source detection and selection and the computation of the sky
coverage are described in detail in Cagnoni, Della Ceca and Maccacaro
(1998) and Della Ceca et al. (1999).

The 300 GIS2 images adequate for this project have been searched for
sources with a signal-to-noise (S/N) ratio greater than 4.0 (a more
restrictive criterion than that adopted in Cagnoni et al., (1998) where
a S/N $\geq$ 3.5 was used).  A sample of 189 serendipitous sources with
fluxes in the range $\sim 1 \times 10^{-13} - \sim 7 \times 10^{-12}$
erg cm$^{-2}$ s$^{-1}$, found over a total area of sky of $\sim 71$
deg$^2$, has been defined.  Full details on this sample will be
reported in Della Ceca at al., (2000).

\section{The 2-10 keV LogN($>$S)$-$LogS}

In Figure 1 we show a parametric (solid line) and a non parametric (solid
histogram) representation of the number-flux relationship obtained 
using the new ASCA HSS sample of 189 sources.

Also shown in Figure 1 (cross at $\sim 3\times 10^{-11}$ ergs cm$^{-2}$
s$^{-1}$) is the surface density of the extragalactic population in the
Piccinotti et al., (1982) HEAO 1 A-2 sample 
(as corrected by Comastri et al., 1995)
and the surface density of
X-ray sources as determined by Kondo (1991) using a small
sample of 11 sources extracted from the Ginga High Galactic Latitude 
survey (filled triangle at  $\sim 8\times 10^{-12}$ ergs cm$^{-2}$
s$^{-1}$).
The surface densities represented by the filled dots at 
$\sim 1.2\times 10^{-13}$, 
$\sim 1.8\times 10^{-13}$, and
$\sim 3.0\times 10^{-13}$ ergs cm$^{-2}$ s$^{-1}$ 
are the results from the ASCA Large Sky Survey (Ueda et al., 1999); 
the filled dot at 
$\sim 5\times 10^{-14}$ ergs cm$^{-2}$ s$^{-1}$ 
has been obtained by Georgantopoulos et
al. (1997) using 3 deep ASCA GIS observations; 
the filled dot at 
$\sim 4.0\times 10^{-14}$ ergs cm$^{-2}$ s$^{-1}$ 
has been obtained from Inoue et al. (1996) using data from a 
deep ASCA observation. 
Finally, the filled square at 
$\sim 5.0\times 10^{-14}$ ergs cm$^{-2}$ s$^{-1}$ 
has been obtained by Giommi et al. (1998) using data from the BeppoSAX 
deep surveys. 
As it can be seen, our determination of the number-flux relationship is 
in very good agreement with those obtained from other survey programs.

\begin{figure}
\centerline{\psfig{file=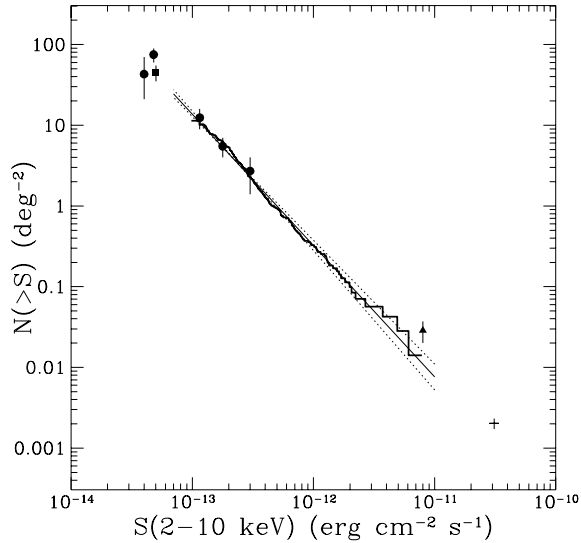, width=8cm}}
\caption[]{The 2-10 keV logN($>$S)-logS. See section 3 for details.}
\end{figure}

The LogN($>$S)-LogS can be described by a power law model N($>$S) = $K
\times S^{-\alpha}$ with best fit value for the slope of $\alpha = 1.63
\pm 0.09$; the dotted lines represent the $\pm 68\%$ confidence
intervals on the slope.  The normalization K is determined by rescaling
the model to the actual number of objects in the sample and, in the
case of the ``best" fit model, is $K=9.65\times 10^{-21}$
deg$^{-2}$.  At the flux limit of the survey ($\sim 7 \times
10^{-14}$ ergs cm$^{-2}$ s$^{-1}$) the total emissivity of the resolved
objects is  $\sim 10$ keV cm$^{-2}$ s$^{-1}$ sr$^{-1}$, i.e.
about 30\% of the 2-10 keV CXB. A
flattening of the number-flux relationship, within a factor of 10 from
our flux limit, is expected in order to avoid saturation.

\section{The 2-10 keV spectral properties of the sources}

To investigate the spectral properties of the sources in the
2.0 - 10.0 keV energy range we defined the Hardness Ratio, 
$HR2 = {H-M \over H+M}$, where H and M are the observed (GIS2 + GIS3)
net counts in the 2.0-4.0 keV and 4.0-10.0 keV energy band
respectively (see Della Ceca et al.,1999 for details).
In Figure 2a, for all sources, we plot the HR2 value versus the GIS2
count rate; we have also reported the flux scale obtained assuming a
count rate to flux conversion factor appropriate for a power law model with
$\alpha_E \sim 0.6$, the median energy spectral index of the sample.
The HR2 values are then compared with those expected from a non
absorbed power-law model with $\alpha_E$ ranging from $-$1.0 to 2.0.
It is worth noting the presence of many sources which seem to be
characterized by a very flat 2-10 keV spectrum with $\alpha_E \leq 0.4$
and of a number of sources with ``inverted" spectra (i.e.  $\alpha_E \leq
0.0$).  

A flattening of the mean spectrum of the sources with decreasing count
rate is clearly evident.  If we divide the sample into two subsamples
(the bright sample is defined by the 60 sources with a count rate
$\geq 4.3\times 10^{-3}$ cts s$^{-1}$, while the faint sample
is defined by the remaining 129 sources), than the fraction of 
sources with $\alpha_E \leq 0.4$ ($\alpha_E \leq 0.0$) is $15 \pm 5 \%$
($8 \pm 4 \%$) in the bright sample and becomes $43 \pm 7 \%$ ($18
\pm 4 \%$) in the faint sample.
These objects with very flat spectra could represent a new population of
very hard serendipitous sources or, alternatively, a population of very
absorbed sources as expected from the CXB synthesis models based on the
AGN Unification Scheme.

\begin{figure}
\psfig{file=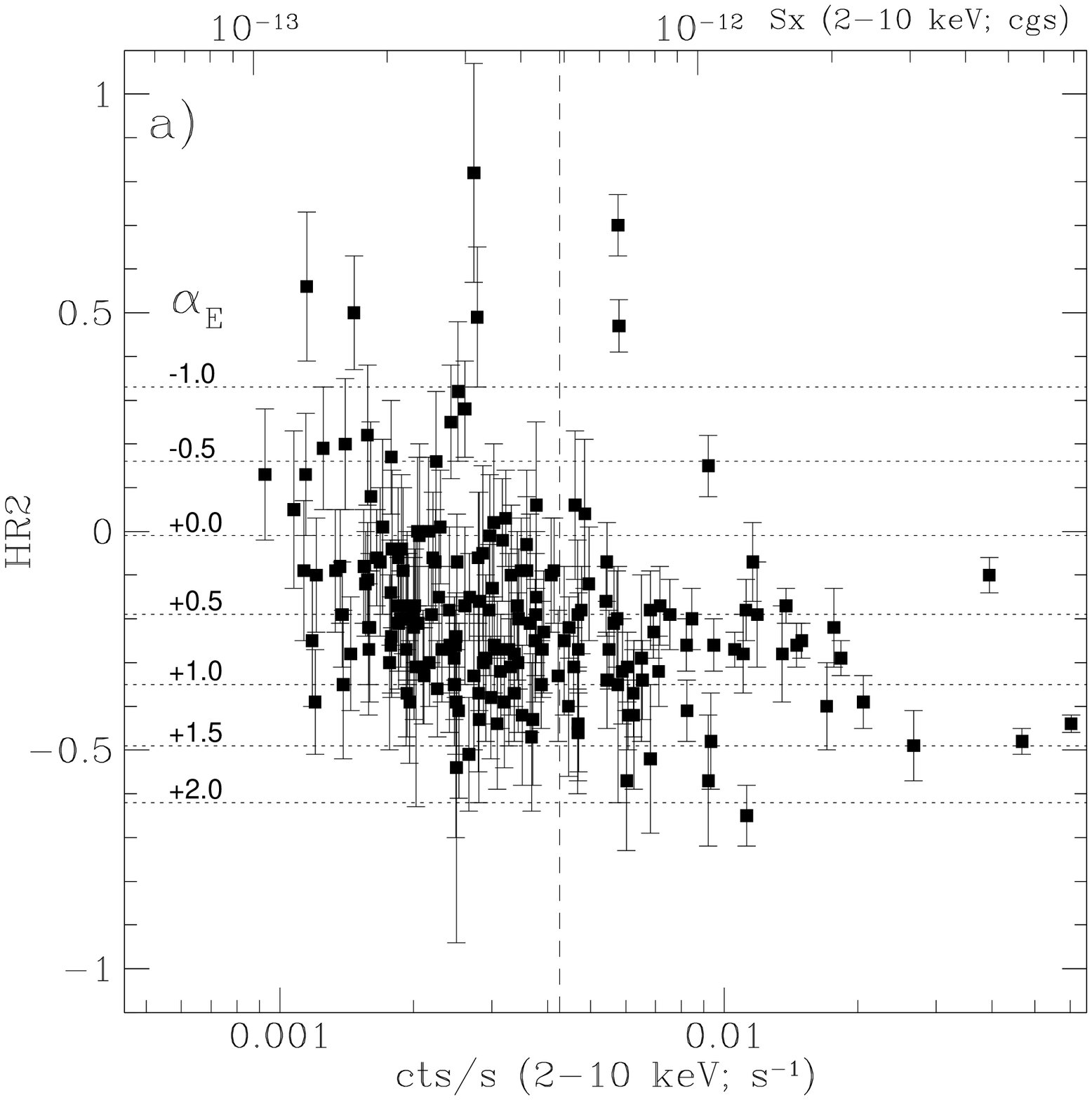, width=6.5cm}
\vskip -6.5truecm
\hskip +6.0truecm
\psfig{file=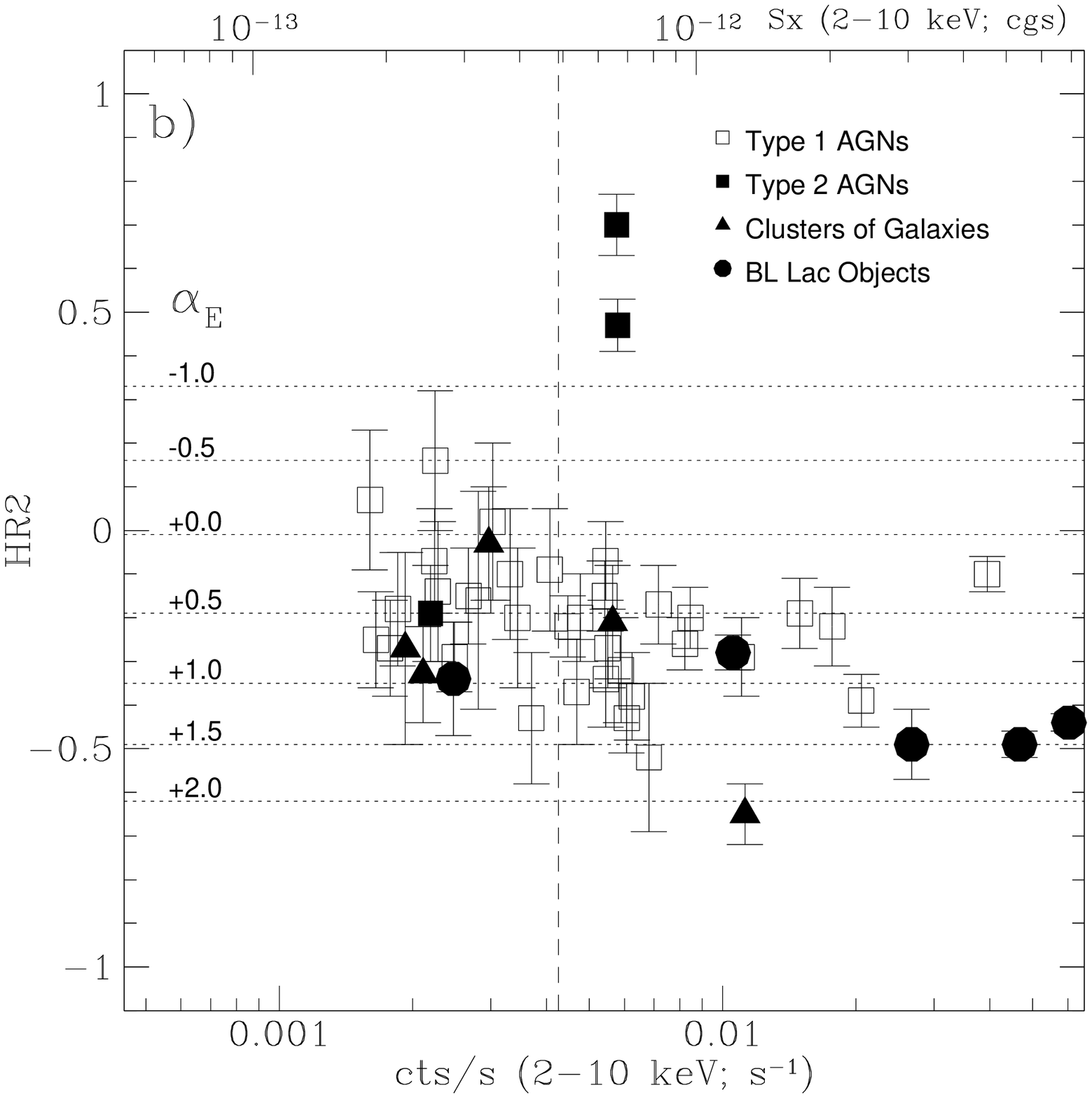, width=6.5cm}
\caption[]{HR2 vs. count rate. Panel a: the complete 
ASCA HSS sample; Panel b: the identified objects}
\end{figure}

\section{The spectroscopically identified sample}

Up to now 47 sources have been spectroscopically identified.
The optical breakdown is the following: 1 star, 5 cluster of galaxies,
5 BL Lac objects, 33 Broad Line Type 1 AGNs and 3 Narrow Line Type 2 
AGNs. However we stress that this
small sample of identified objects is probably not representative of
the whole population.

In Figure 2b we plot the HR2 value versus the GIS2
count rate for this small sample of identified objects.
We note that 2 of the 3 objects classified as Type 2 AGNs 
have an inverted X-ray spectrum in the 2-10 keV band, and 
that some of the Type 1 AGNs seem to have a very flat
($\alpha_E \leq 0.5$) spectrum.


\begin{acknowledgements}
This work received partial financial support from the Italian 
Ministry for University Research (MURST) under grant Cofin98-02-32
and from the Fondazione CARIPLO.
\end{acknowledgements}


\begin{references}

\ref Cagnoni, I., Della Ceca, R., and Maccacaro, T., 1998, Ap.J., 493, 54.

\ref Comastri, A., Setti, G., Zamorani, G., and Hasinger, G., 1995, A\&A, 296, 1.

\ref Della Ceca, R., Castelli, G., Braito, V., Cagnoni, I., and Maccacaro, T., 1999, 
Ap.J., 524, 674.

\ref Della Ceca, R., et al., 2000, in preparation.

\ref  Georgantopoulos, I., et al., 1997, MNRAS, 291, 203.

\ref Giommi, P., et al., 1998, Nuclear Physics B (Proc. Suppl.), 69/1-3, 591.

\ref Inoue, H., Kii, T., Ogasaka, Y., Takahashi, T., and Ueda, Y., 1996, MPE REP. 263, 323. 

\ref  Kondo, H., 1991, Ph.D. Thesis Univ. of Tokyo.

\ref Piccinotti, G., et al., 1982, Ap.J., 253, 485.

\ref Ueda, Y., et al., 1999, Ap.J., 518, 656.



\end{references}
\end{document}